# New Method for Studying the Response of the Slot Antennas in the Presence of HEMP

Atefe Akbari-Bardaskan
Department of Electrical Engineering, Ferdowsi University of Mashhad, Mashhad, Iran

*Abstract*—Electromagnetic Interference that is created by any methods can have an important effect on electronic and telecommunication devices. Thus, the effect of electromagnetic waves on communication systems is particularly important. Nowadays, the slot antennas are an inseparable part of many communication devices. In this paper, two methods have been introduced to evaluate the effect of interference wave on slot antennas. The first method is a classic technique based on Maxwell's equations that studied the impact of electromagnetic waves on slot antennas by using of the radiation integrals. In the second method, the slots, by using the Babinet's principle are modeled with the linear antenna. Then, using linear antenna circuit and then, the equivalent circuit antennas at the receiver mode is used. Eventually, the effect of interfering waves, by calculating the voltage induced in the antenna port, is investigated. Simulation results are given to verify the usefulness of the two introduced methods. The results clearly show the reasonable outcome of the proposed methods in study of electromagnetic interference compared to the simulation results.

## I. INTRODUCTION

Single antennas, slot antennas, and phased array antennas are widely used in electrical, electronic, and communications systems [1-10]. High-power electromagnetic pulses are usually created by high-altitude nuclear explosions [11-12]. These waves due to have high power pass from many objects easily and influence on electronic and telecommunication systems, such as non-uniform guiding structures, microstrip lines, substrate integrated waveguides, etc. [13-19]. The investigation of electromagnetic compatibility (EMC) and electromagnetic interference (EMI) is an inseparable part of projects have been carried out, particularly in electronic equipment. This is mainly due to create an environment free of noise to provide the basis for radio communications. So, study of the impact of electromagnetic waves on the communication devices is an interesting problem to many researchers. In recent decades, slot antennas have been used in many different fields of military and industry applications such as marine radar. Slot antennas are used usually in the frequency range from 300MHz to 25GHz.

Coupling energy of EMP through an aperture in a perfectly conducting thin plane has been the subject of many researches. In 1976, integral-differential equations for the problem of coupling through an aperture of general shape have been presented by Butler [20]. Well as Butler, Harrington and Mautz [21] reduced system to the linear equations by using equivalence principle, the continuity of the tangential field components and Method of Moment. In [22], a hybrid numerical technique has been presented to calculate the scattering and transmission properties of a slot in a thick conducting plane.

Unlike the accuracy of these methods that have been introduced, they are time consuming, need large amount of memory and complicated formula has been used for this purpose. This article focuses on providing the simple and functional strategy to determine the response of the slot antennas in the presence of high-altitude electromagnetic pulses. In this paper, two efficient methods with reasonable accuracy are used to solve the problem of passing the electromagnetic waves through an aperture on a thin conducting plane.

In the first method and according to the equivalence principle, the electrical and magnetic current density on the slot are determined and then by using the free space Green's function, the radiated electromagnetic fields due to incident waves on slotted conductor plate, can be calculated. In fact, in this method by using the Maxwell's equations, electric and magnetic fields can be obtained from the auxiliary vector potentials. In the second method, the slot on the conductor plate by using the Babinet's principle is modeled with the linear dipole antenna. Then, the equivalent circuit for dipole antenna at the receiver mode is applied and by determining the voltage induced in the antenna port by use of circuit theory, the effect of incident electromagnetic waves can be observed. Finally, several examples to check the accuracy of the proposed method will be considered.

## II. HEMP ENVIRONMENT

The electromagnetic environment generated by high altitude nuclear explosion is usually divided into three components: early time (E1 HEMP), intermediate (E2 HEMP) and late time (E3 HEMP) [23]. Because of the short front time of E1 HEMP and its great field magnitude, it causes the strongest effects. The popular waveform of E1 HEMP described by (1):

$$E_i = E_0 k_0 \left( e^{-at} - e^{-bt} \right); \quad t>0 \qquad (1)$$

Where $E_0$=50 kV/m, $k_0$=1.3, a=4×10$^7$, and b=6×10$^8$. Figure (1) shows the $E_1$ HEMP waveform.

As shown in Figure (2), for an aperture on an infinite (planar or flat), perfectly conducting and vanishingly thin plane, due to the tangential component of electric field of incident wave, an equivalent magnetic current is produced over the aperture region that can be obtained as following:

$$M_s = -\hat{n} \times E_a \quad (2)$$

Where $n$ the normal unit vector and $E_a$ is the tangential component of the electric field over the aperture. According to the image theory, the conducting plane removed and the magnetic current density can be replaced by an imaginary or equivalent source $M_s$. So, the problem of Figure 2a reduces to that of Fig. 2b.

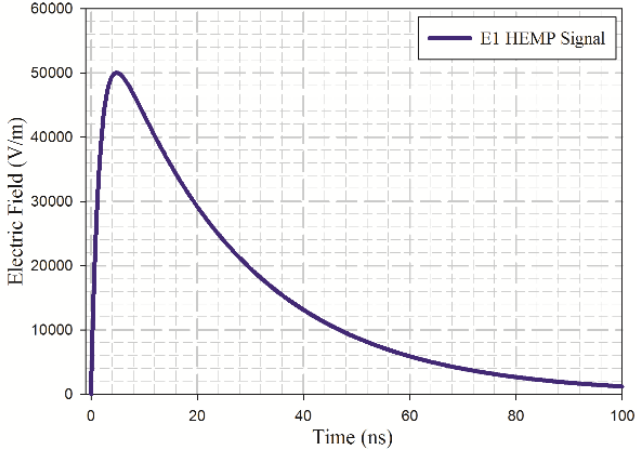

Fig. 1: The E1 HEMP waveform.

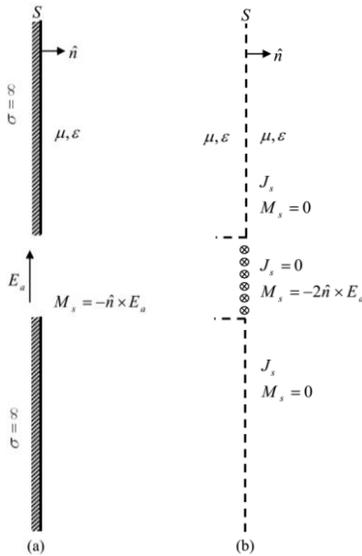

Fig. 2: Equivalent model for an aperture in the planar conducting screen of infinite extent.

Now, for calculating the radiation fields of the equivalent source $M_s$, electric vector potential $F$ can be calculated as following:

$$F(x,y,z) = \frac{\varepsilon}{4\pi} \iiint_{v'} M_s(x',y',z') \frac{e^{-j\beta R}}{R} dv' \quad (3)$$

Where (x, y, z) represent the observation point, (x', y', z') represent the source coordinate and R is the distance from any point on the source to the observation point.

Since obtaining a closed form solution for (3) is usually difficult, some approximations are used to simplify the formulation of the radiated field by an antenna to yield close form solution [24].

The most practical approximation for radiated field by the antenna is far-field approximation. So for a 2-dimensional source in xz plane, far-field approximation is as following:

$$R = r - r'\cos\psi = r - (x'\hat{a}_x + z'\hat{a}_z) \cdot$$
$$(\sin\theta\cos\phi\,\hat{a}_x + \sin\theta\sin\phi\,\hat{a}_y + \cos\theta\,\hat{a}_z) \quad (4a)$$
$$= x'\sin\theta\cos\phi + z'\cos\theta$$

$$R \approx r \quad (4b)$$

According to (4), (3) can be rewritten as follows:

$$F = \frac{\varepsilon}{4\pi} \frac{e^{-j\beta r}}{r} \iint_{s'} M_s e^{-j\beta(x'\sin\theta\cos\phi + z'\cos\theta)} ds' \quad (5)$$

By solving equation (5), radiated electric and magnetic fields can be obtained from equations (6) and (7) respectively [24]:

$$E_F = \frac{-1}{\varepsilon} \nabla \times F \quad (6)$$

$$H_F = -j\omega F + \frac{\nabla(\nabla \cdot F)}{j\omega\mu\varepsilon} \quad (7)$$

For a TEM wave which propagates in $y$-direction and its electric field is in $z$-direction the equivalent magnetic current is in $x$-direction. Therefore the vector potential $F$ just has an $x$-component. It should be noted that, the equations have been presented, are just valid for single frequency signals. Since the HEMP signal is not a single frequency wave, we have to break it into single frequency signal by using Fourier's series of it. Even expansion of HEMP is as follows:

$$E = \begin{cases} E(t) & t>0 \\ E(-t) & t<0 \end{cases} \quad (8)$$

So by considering the first 100 components of Fourier's series, the HEMP signal can be modeled correctly, as shown in Fig. 3. Now, the total radiation field of HEMP can be calculated by sum of the radiated electric field of each term of Fourier's series.

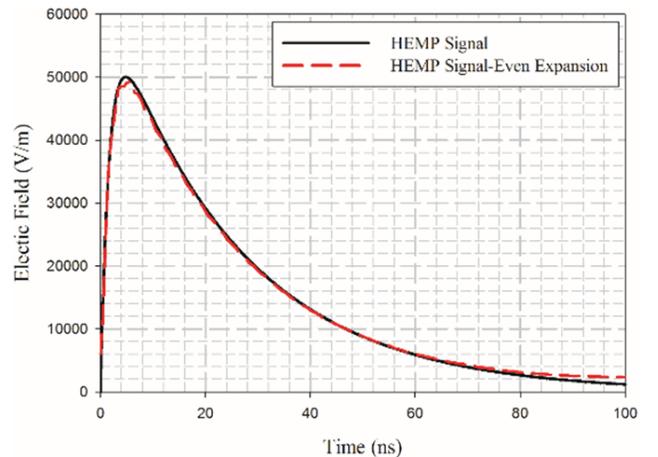

Fig. 3: The HEMP signal and its Fourier's even expansion.

In this method, by using of the Babinet's principle [24], it can use a thin dipole antenna equation for solving the problem of electromagnetic transmission through a slot. According to [23], the equivalent circuit of an antenna in receiving mode is shown in Fig. 4.

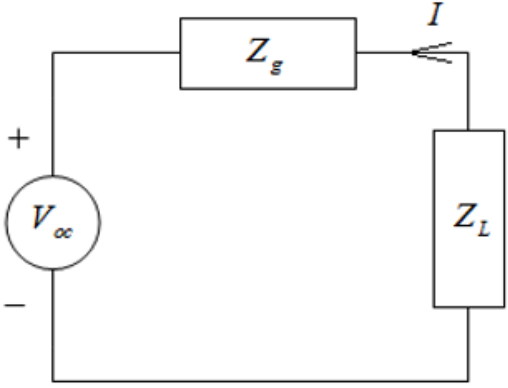

Fig. 4: The equivalent circuit of an antenna in receiving mode.

In this figure, $V_{oc}$ is induced open circuit voltage of the antenna, $I$ is induced current in the antenna and $Z_{in}$ and $Z_L$ are input impedance of the antenna and load impedance, respectively. $V_{oc}$ can be calculated by (9):

$$V_{oc} = -E_i \cdot \ell_e(\theta) \qquad (9)$$

Where $E_i$ is the incident electric field and $\ell_e$ is effective length vector of antenna. For a thin dipole, $\ell_e$ can be obtained from radiated electric field of the antenna, by (10), $\ell_e$ is.

$$\ell_e(\theta) = \frac{2}{\beta} \frac{\cos\left(\frac{\pi}{2}\cos\theta\right)}{\sin\theta}(\cos\theta \hat{a}_x - \sin\theta \hat{a}_z) \qquad (10)$$

As shown in Fig. 5, if the incident wave is orthogonal to the antenna axis, the angle $\theta$ in (10) becomes $\pi/2$. So for a normal incidence wave, the induced open circuit voltage will be as following:

$$V_{oc} = E_i \cdot \ell_e = (2E_i/\beta) \qquad (11)$$

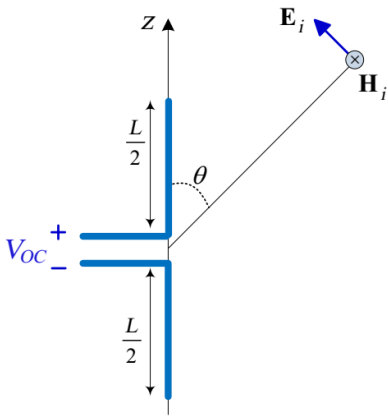

Fig. 5: A linear antenna in receiving mode.

For determining $Z_{in}$, we can use Babinet's principle. Babinet's principle states that the relation between the slot impedance, $Z_s$ and the impedance of its complementary structure, $Z_d$ is as following equation:

$$Z_s Z_d = \frac{\eta_0^2}{4} \qquad (12)$$

Where $\eta_0$ is the free space characteristic impedance. Now by writing KVL in the loop of the circuit in Figure 4, the induced current due to an incident wave can be determined. Usually, for real values of $Z_L$ and $Z_g$, time variable receiving voltage can be calculated as:

$$V_{oc} = -(Z_g + Z_L)I \Rightarrow I = \frac{-V_{oc}}{(Z_g + Z_L)} \qquad (13)$$

If both load and source impedance are complex values, the equivalent circuit can be solved with Laplace transform method and by taking the inverse Laplace transform, time variable receiving current will be calculated. It should be noted that, as we mentioned before, the equations have been presented, are just valid for single frequency signals. Since the HEMP signal and the induced current are not single frequency signals, we have to break the current into single frequency signal by using Fourier's series of it. Even expansion of the current, $I'(t)$ is as follows:

$$I'(t) = \begin{cases} I(t) & t>0 \\ I(-t) & t<0 \end{cases} \qquad (14)$$

So by writing first 200 components of Fourier's series, the current signal can be modeled with reasonable accuracy. As a result, the total radiated field is respect to the sum of the radiated fields of each component of Fourier's series.

## III. NUMERICAL AND SIMULATION RESULTS

In CST Studio Suite 2015, an aperture in thin conducting plane, several cases is considered for the validation of the two proposed methods. Although the two proposed method are based on this assumption that the conducting plane is vanishingly thin with infinite extend, applying this assumption is not available in CST due to the limitation of the memory of systems. Therefore, a metal sheet with a slot of at least $\lambda$ perimeter may transmit considerable energy. In the simulation environment, HEMP signal, Figure 1 applied as the excitation signal, in the way that the propagating direction is $y$, the electric field is in $z$-direction and the magnetic field is in $x$-direction.

For determining the far-field radiation in CST software, the probe is set in the distance $r$ from origin. As shown in Fig. 6, the proposed method can predict radiated field for the slot which the length of the slot is not too big in proportion to the width of it, with good approximation. First by locating a resistance in the slot, the induced current due to the HEMP is measured.

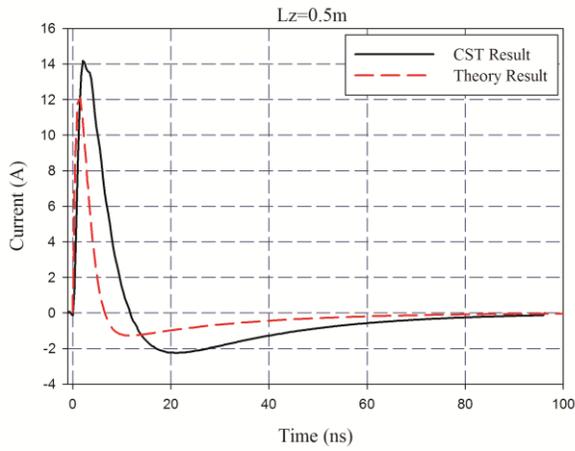

Fig. 6: The Induced current in the resistance, calculated by theory and simulation.

The radiated field of this current distribution calculated by (13), in comparison with simulation results are shown in the following Figures. The graphs show that this method is suitable for narrow slots.

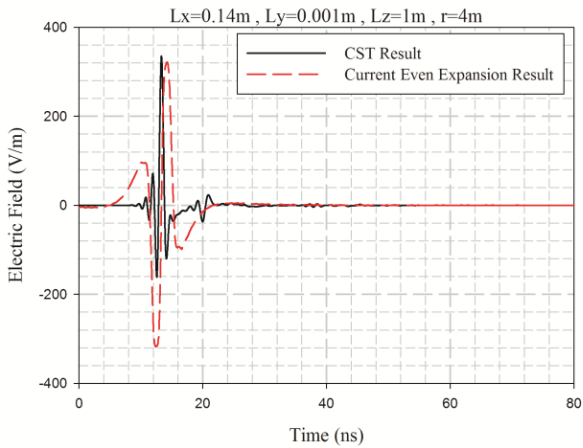

Fig. 7: The Far-field radiation calculated by the second method and CST results, for the slot with $L_x$=0.14m, $L_y$=0.001m, $r$=4m.

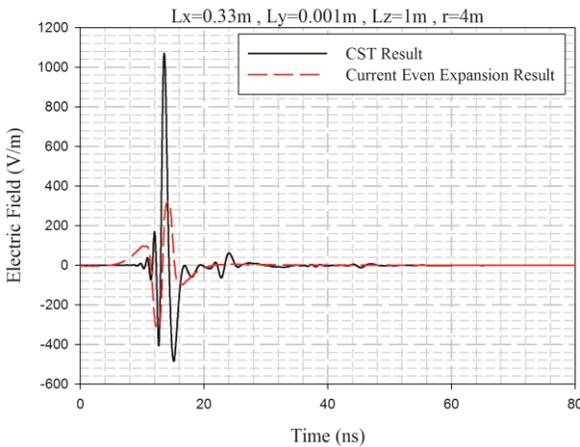

Fig. 8: The Far-field radiation calculated by the second method and CST results, for the slot with $L_x$=0.33m, $L_y$=0.001m, $r$=4m.

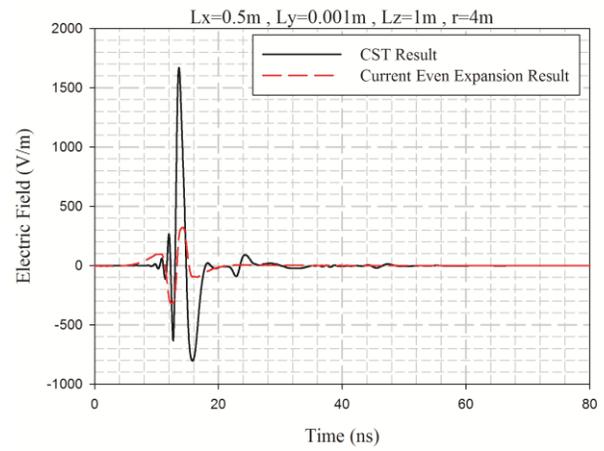

Fig. 9: The Far-field radiation calculated by the second method and CST results, for the slot with $L_x$=0.5m, $L_y$=0.001m, $r$=4m.

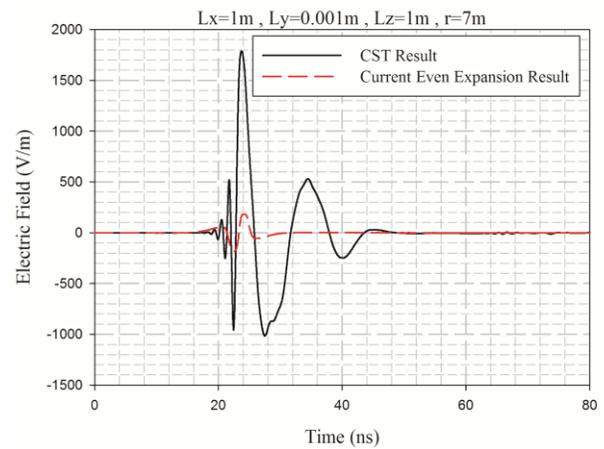

Fig. 10: The Far-field radiation calculated by the second method and CST results, for the slot with $L_x$=1m, $L_y$=0.001m, $r$=4m.

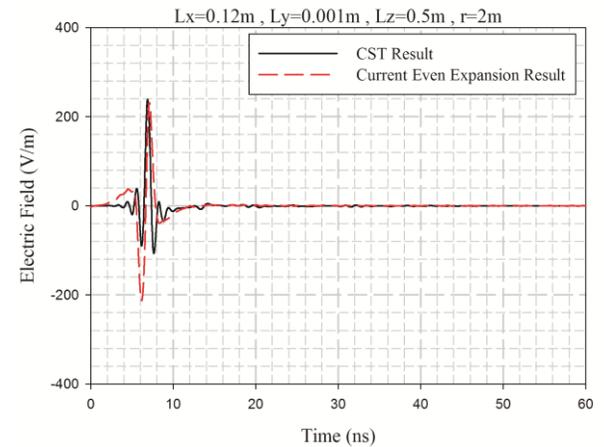

Fig. 11: The Far-field radiation calculated by the second method and CST results, for the slot with $L_x$=0.12m, $L_y$=0.001m, $L_z$=0.5m $r$=2m.

Note that both methods are obtained under some special circumstances, such as infinite extension of the plane which the slot is created in. Moreover, in the second method the basic assumption for using current distribution given by (13) and Babinet's principle is $L \gg w$. Another important factor that decreases the accuracy of the two methods is that in the simulation environment diffraction and scattering of waves are considered, whereas in the proposed methods, these are ignored. Consideration of them will add more complicated calculation to the method.

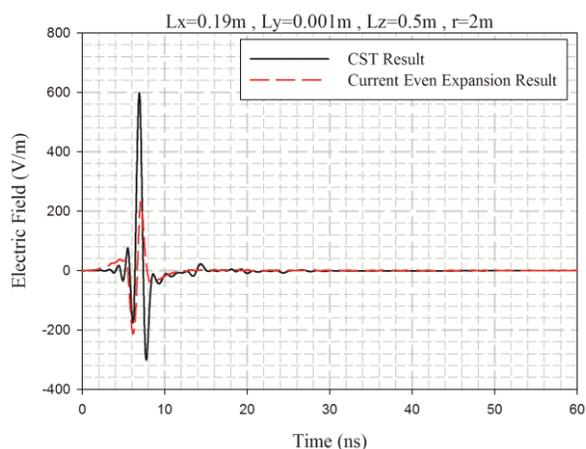

Fig. 12: The Far-field radiation calculated by the second method and CST results, for the slot with $L_x$=0.19m, $L_y$=0.001m, $L_z$=0.5m $r$=2m.

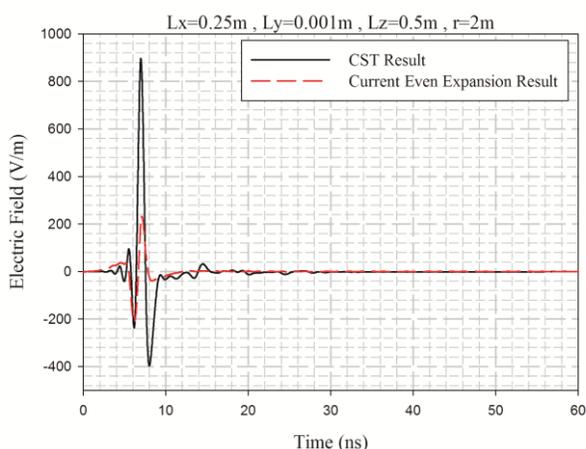

Fig. 13: The Far-field radiation calculated by the second method and CST results, for the slot with $L_x$=0.25m, $L_y$=0.001m, $L_z$=0.5m $r$=2m.

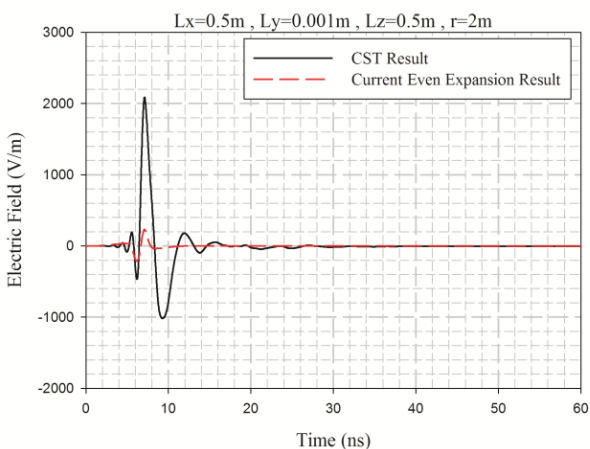

Fig. 14: The Far-field radiation calculated by the second method and CST results, for the slot with $L_x$=0.5m, $L_y$=0.001m, $L_z$=0.5m $r$=2m.

## IV. CONCLUSION

High-power electromagnetic waves, when applied to electrical devices, create voltage and current with high amplitude in the circuit. This phenomenon can disable the equipment completely or interfere in its performance. Such waves are called to high altitude electromagnetic pulse or HEMP. Study and analyze the impact of high-power electromagnetic waves on electronic and communication devices, is an important issue that has attracted the attention of EMC and EMI engineers. Slot antenna due to features such as easy fabrication, low cost, wide scan range and etc, has been widely used in military and commercial industries. In the past, large range of analytical techniques to study the effects of electromagnetic waves on antennas have been used that many of these methods were complex and have low accuracy. Today, commercial software is used for this purpose to reduce the cost. However, the use of full-wave simulators, are time consuming. Thus, a simple method with high precision saves on both costs and time considerably.

This paper concentrates on offering the simple and purposeful procedure to determine the response of the slot antennas in the presence of the HEMP. For this purpose, two efficient methods with rational exactness are introduced. In the first method by use of the equivalence principle, the electrical and magnetic current density on the slot is calculated and by use of the spherical Green's function, the auxiliary vector potentials are determined. Then, the radiated field from the auxiliary vector potentials can be calculated. In the second method and according to the Babinet's principle, slot modeled with dipole antenna. Then, the equivalent circuit for dipole antenna at the receiver mode is utilized to obtain the voltage induced in the antenna port. Finally, several examples for both two methods are considered. The results clearly show the proposed methods have a logical accuracy respect to the simulation results.

REFERENCES

[1] M. G. H. Alijani, S. Sheikh and A. Kishk, "Development of Closed-Form Formula for Quick Estimation of Antenna Factor," *2021 15th European Conference on Antennas and Propagation (EuCAP), Dusseldorf, Germany*, 2021, pp. 1-5, doi: 10.23919/EuCAP51087.2021.9411008.
[2] K. Marák, J. Kracek and S. Bilicz, "Antenna Array Pattern Synthesis Using an Iterative Method," *IEEE Transactions on Magnetics*, vol. 56, no. 2, pp. 1-4, Feb. 2020, Art no. 7508304, doi: 10.1109/TMAG.2019.2952809.
[3] A. Tamandani and M. G. H. Alijani, "Development of An Analytical Method for Pattern Synthesizing of Linear and Planar Arrays with Optimal Parameters," *AEU-International Journal of Electronics and Communications*, vol. 146, pp. 1-7, March 2022, doi: 10.1016/j.aeue.2022.154135.
[4] H. Kähkönen, J. Ala-Laurinaho and V. Viikari, "A Modular Dual-Polarized Ka-Band Vivaldi Antenna Array," *IEEE Access*, vol. 10, pp. 36362-36372, 2022, doi: 10.1109/ACCESS.2022.3164201.
[5] M. G. H. Alijani and M. H. Neshati, "Development a New Technique Based on Least Square Method to Synthesize the Pattern of Equally Space Linear Arrays," *International Journal of Engineering, Transactions B: Applications*, vol. 32, no. 11, pp. 1620-1626, November 2019, doi: 10.5829/ije.2019.32.11b.13.
[6] A. Tamandani and M. G. H. Alijani, "Development of a new method for pattern synthesizing of the concentric ring arrays with minimum number of rings," *AEU-International Journal of Electronics and Communications*, vol. 152, pp. 1-7, July 2022, doi: 10.1016/j.aeue.2022.154262.
[7] C. Z. Feng, W. T. Li, C. Cui, Y. Q. Hei, J. C. Mou and X. W. Shi, "An Efficient and Universal Static and Dynamic Convex Optimization for Array Synthesis," *IEEE Antennas and Wireless Propagation Letters*, vol. 21, no. 10, pp. 2060-2064, Oct. 2022, doi: 10.1109/LAWP.2022.3190421.
[8] M. G. H. Alijani and M. H. Neshati, "Development of a New Method for Pattern Synthesizing of Linear and Planar Arrays Using Legendre

Transform With Minimum Number of Elements," *IEEE Transactions on Antennas and Propagation*, vol. 70, no. 4, pp. 2779-2789, April 2022, doi: 10.1109/TAP.2021.3137200.

[9] P. Angeletti, L. Berretti, S. Maddio, G. Pelosi, S. Selleri and G. Toso, "Phase-Only Synthesis for Large Planar Arrays via Zernike Polynomials and Invasive Weed Optimization," *IEEE Transactions on Antennas and Propagation*, vol. 70, no. 3, pp. 1954-1964, March 2022, doi: 10.1109/TAP.2021.3119113.

[10] M. Sharifi, M. Boozari, M. G. Alijani and M. H. Neshati, "Development a New Algorithm to Reduce SLL of an Equally Spaced Linear Array," *Electrical Engineering (ICEE), Iranian Conference on, Mashhad, Iran*, 2018, pp. 554-557, doi: 10.1109/ICEE.2018.8472414.

[11] W. A. Radasky, C. E. Baum and M. W. Wik, "Introduction to the special issue on high-power electromagnetics (HPEM) and intentional electromagnetic interference (IEMI)," *IEEE Transactions on Electromagnetic Compatibility*, vol. 46, no. 3, pp. 314-321, Aug. 2004, doi: 10.1109/TEMC.2004.831899.

[12] W. D. Prather, C. E. Baum, R. J. Torres, F. Sabath and D. Nitsch, "Survey of worldwide high-power wideband capabilities," *IEEE Transactions on Electromagnetic Compatibility*, vol. 46, no. 3, pp. 335-344, Aug. 2004, doi: 10.1109/TEMC.2004.831826.

[13] M. G. H. Alijani, S. Sheikh, A. A. Kishk, "Analytical method for single and coupled nonuniform guiding structures," *AEU-International Journal of Electronics and Communications*, vol. 162, pp. 1-8, April 2023, doi: 10.1016/j.aeue.2023.154590.

[14] H. Karimian-Sarakhs, M. H. Neshati, M. G. H. Alijani, "Development of an analytical method to determine the propagation characteristics of microstrip line on artificial perforated substrates," *AEU-International Journal of Electronics and Communications*, vol. 140, pp. 1-8, October 2021, doi: 10.1016/j.aeue.2021.153951.

[15] Y. Yang, J. Tan, D. Sheng, and G. Yang, "Response characteristics and protection techniques of monopole on conductive plane exposed to electromagnetic pulse," *High Power Laser and Particle Beams*, vol. 20, no. 4, 2008. doi:10.3788/HPLPB20152704.41013.

[16] A. Tamandani and M. G. H. Alijani, "Analysis of a nonuniform guiding structure by the adaptive finite-difference and singular value decomposition methods," *ETRI journal*, vol. 45, no. 24, pp. 1-9, March 2023, doi: 10.4218/etrij.2022-0076.

[17] N. Yasi, M. Boozari, M. G. H. Alijani, M. H. Neshati, "A new method for analyzing non-uniform guiding structures," *Scientific Reports*, vol. 13, pp. 1-11, February 2023, doi: 10.1038/s41598-023-30145-6.

[18] W. A. Radasky and R. Hoad, "Recent Developments in High Power EM (HPEM) Standards With Emphasis on High Altitude Electromagnetic Pulse (HEMP) and Intentional Electromagnetic Interference (IEMI)," *IEEE Letters on Electromagnetic Compatibility Practice and Applications*, vol. 2, no. 3, pp. 62-66, Sept. 2020, doi: 10.1109/LEMCPA.2020.3009236.

[19] M. Alijani-Ghadikolae and M. H. Neshati, "Developing an accurate and simple dispersion analysis of TE10 mode of substrate integrated waveguides," *2013 21st Iranian Conference on Electrical Engineering (ICEE)*, Mashhad, Iran, 2013, pp. 1-4, doi: 10.1109/IranianCEE.2013.6599879.

[20] C. M. Butler and K. R. Umashankar, "Electromagnetic penetration through an aperture in an infinite, planar screen separating two half spaces of different electromagnetic properties," *Radio Science*, vol. 11, no. 7, pp. 611-619, July 1976, doi: 10.1029/RS011i007p00611.

[21] R. F. Harrington and J. R. Mautz, "Electromagnetic transmission through an aperture in a *conducting plane*," *Archiv Elektronik und Uebertragungstechnik*, vol. 31, pp. 81 -87, February 1977, doi: 1977ArElU..31...81H.

[22] Y. Rahmat-Samii and R. Mittra, "Electromagnetic coupling through small apertures in a conducting screen," *IEEE Transactions on Antennas and Propagation*, vol. 25, no. 2, pp. 180-187, March 1977, doi: 10.1109/TAP.1977.1141554.

[23] M. Boozari, M. G. H. Alijani, "Investigation on Transient Response of Linear Dipole Antennas in the Presence of HEMP Based on Equivalent Circuit," *Progress In Electromagnetics Research Letters*, vol. 66, pp. 39-43, 2017, doi: 10.2528/PIERL16123006.

[24] C. A. Balanis, *Antenna Theory: Analysis and Design*. Hoboken, Nj, USA: Wiley, 2016.